\documentclass[useAMS,usenatbib]{mn2e}
\usepackage{graphics}
\usepackage{graphicx}
\usepackage{multirow}
\usepackage{amssymb}
\usepackage{color}
\newcommand{\be}{\begin{equation}}
\newcommand{\ee}{\end{equation}}
\def \vs {{\it vs.}\ }

\def \etal {{\it et al.}\ }
\def \LCDM {$\Lambda$CDM}

\newcommand{\hmsun}{{\,\rm h^{-1}M}_\odot}
\newcommand{\hmpc}{{\,\rm h^{-1}Mpc}}
\def\bg{{\bf g}}
\def\br{{\bf r}}

\def\bff{{\bf f}}
\def\bv{{\bf v}}

\newcommand\apjl{ApJ}%
\newcommand\apj{ApJ}%
\newcommand\mnras{MNRAS}%
\newcommand\aj{AJ}%

\begin{document}

\title{Constrained simulations of the local universe:  I. Mass and motion in the Local Volume}
\author[Martinez-Vaquero et al.]
{\parbox[t]\textwidth{Luis A. Martinez-Vaquero$^1$, Gustavo Yepes$^1$ and Yehuda Hoffman$^2$}
\vspace*{6pt} \\
$^1$Grupo de Astrof\'{\i}sica, 
Universidad Aut\'onoma de Madrid,
Madrid E-280049, Spain 
\\
$^2$Racah Institute of Physics, 
Hebrew University, 
Jerusalem 91904, Israel 
\\
}
\date{\today}

\maketitle

\begin{abstract}
It has been  recently claimed that there is no correlation between the
distribution of galaxies and their peculiar velocities within the  Local Volume (LV), namely a sphere of $R=7\hmpc$ around the Local   Group (LG). It has been then stated that this   implies  that either locally dark matter is not distributed in the same way as  luminous matter, or peculiar velocities are not due to fluctuations  in mass.  To test  that statement a  set of  constrained N-body  cosmological simulations, designed   to reproduce the main  observed large scale   structure,  have been analyzed.  
The   simulations  were  performed within the flat-$\Lambda$, open and flat matter only CDM cosmogonies. Two unconstrained   simulations of the flat-$\Lambda$ and open CDM models were performed for   comparison.  LG-like objects have been selected so as to mimic the
real LG environment.    The local    gravitational field due to all halos found within each LV is  compared with the exact gravitational field  induced by all matter in  the simulation.  
We  conclude that there is no  correlation between the   exact and the local gravitational field obtained by pairwise  newtonian forces between halos. Moreover, the local
gravitational  field is uncorrelated with the peculiar velocities of halos.  The  exact gravitational field has a linear correlation with peculiar velocities but the proportionality constant relating the  velocity with gravitational field falls below the prediction of the linear theory.  Upon considering all matter inside the LVs, the exact and local gravitational accelerations show a much better correlation, but with a considerable scatter   independent on the cosmological models.    The main conclusion is   that the  lack of  correlation between the       local gravitation and  the peculiar velocity fields  around LG-like objects is
naturally expected in the CDM  cosmologies.
\end{abstract}


\begin{figure}
\resizebox{\linewidth}{!}{\includegraphics{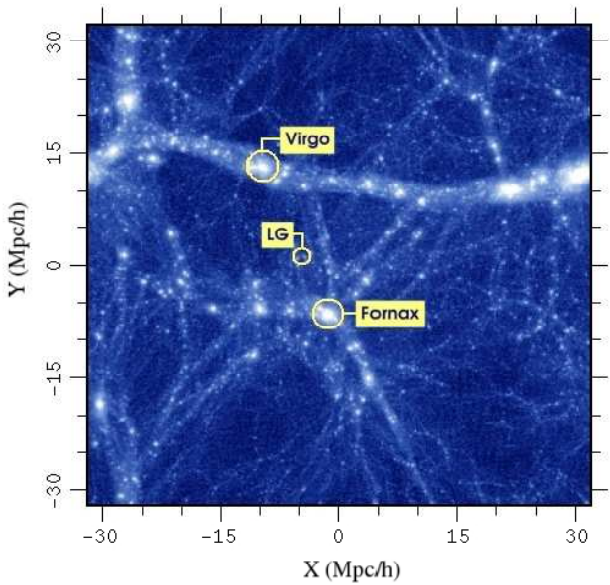}}
\caption {Projected dark matter distribution for the \LCDM\ constrained
simulation. The image shows a projected slice of  10 h$^{-1}$ Mpc thick
across the box centre in  supergalactic coordinates X and
Y. The Local Supercluster is the filamente crossing the box
horizonally. The position of the Local Group and the Virgo and Fornax clusters
 are shown. }
\label{boxL}
\end{figure}

\section{Introduction}

A key ingredient of the standard model of cosmology is that the large
scale structure (LSS) has emerged out of an otherwise a homogenous and
isotropic universe {\it via} gravitational instability
(e.g. \cite{pee80}).
One of the main consequences of gravitational instability is that the
growth of structure induces a non-vanishing velocity field.  The
standard model of cosmology relates the large scale mass density and
peculiar velocity fields. It further assumes that the observed galaxy
distribution is closely related with the underlying matter density
field, even if this relation is biased in some yet unknown way.  In an
interesting recent paper \cite{whi05} (W05) has 
challenged 
these basic ideas of the standard model by testing them against a study
of the galaxy distribution and their peculiar velocities in the Local
Volume (LV), defined as the sphere of radius $R=7 \hmpc$ (where $h$ is the
Hubble's constant in units of $100\, 
\mbox{\rm km s}^{-1} \mbox{\rm Mpc}^{-1}$) centred on
the Local Group (LG). W05's main conclusion is that {\em 'Either dark matter
is not distributed in the same way as luminous matter in this region,
or peculiar velocities are not due to fluctuations in mass.'}
 This is a
very important result and if it 
was
true it would
 put the standard model of cosmology on very shaky foundations. 
The main goal of our  paper is to
examine W05 methodology and claims 
by  analyzing the kinematics and dynamics of simulated LV
sytems in different cosmological models.   As we will show in this
paper,  our numerical results 
do not support
W05's claims.

W05 carried out an analysis of the distribution and peculiar velocities
of 149 galaxies in the LV. Using the high-quality data of these
galaxies W05 has mapped the mass distribution within the LV, assuming
it is traced by the galaxies, and calculated the gravitational field
within the LV.  The gravitational field has been calculated by summing
over the pairwise Newtonian interaction for each galaxy and by
weighting the galaxies by their K and B luminosity. W05's working
assumption is that the peculiar velocity of galaxy should be aligned
with the gravitational field it experiences and that the amplitude of
the peculiar velocity and gravity fields should be linearly connected.
The notion that 'the peculiar velocity field is linear with the
gravitational field' is strictly valid in the linear regime of the
gravitational instability in an expending universe. If such a linear
relation had been confirmed by the high-quality data that the LV
provides this would have validated the idea that the LSS is indeed induced
by gravitational instability.  No clear correlation was  found
between the velocity and local gravitational  fields. This  has
led W05 to conclude that, at least in the LV, 
structure has not formed by means of the gravitational instability.

The above conclusion should be tested carefully.  The main goal of the
present paper is to apply W05 analysis to a set of N-body constrained
simulations (CSs) of the local universe. The CSs are designed to
reproduce the gross features of the nearby universe, namely the main
players of the nearby LSS, such as the Local Supercluster (LSC) and the
Virgo, are imprinted onto the simulations \citep{kra02,mat02,kly04,dol05,hof06}.
 This is achieved by setting the initial conditions of
the simulations as constrained realizations of Gaussian fields, where
actual observational data is used as constraints. The CSs provide an
almost optimal laboratory for testing W05 algorithm, as they closely
mimic the dynamics of the LV. CSs have been run in the framework of the
benchmark flat-$\Lambda$ CDM model as well as  in the open CDM (OCDM) and the
so-called flat matter only standard CDM (SCDM) models. In addition, W05
procedure has been tested against standard, non-constrained,
simulations of the same models.

This is a first in a series of papers that focuses on studying the
nearby universe by means of N-body and hydrodynamical CSs. In
particular issues concerning the coldness of the local flow, the mass
distribution in the LV, the mass accretion history of the LG and the
future of the nearby structure in a dark energy dominated universe are
to be addressed. The highlight of this project will be the full galaxy
formation high resolution simulation of the LG.
Our choice of the different models is dictated by our plan to study the nature of the velocity 
field in the LV in the CDM cosmogony and in particular its dependence on the dark energy
and the dark matter. In this regard the SCDM model is considered to be an extreme case and is 
taken for reference. Forthcoming papers in this series will focus on the problem of the coldness of the local Hubble flow.

The structure of the paper is as follows: 
a brief description of the simulations is given in
\S~\ref{sec:sim}. 
 The selection of LG-like objects is described in \S~\ref{sec:obj} and the
analysis itself in \S~\ref{sec:analysis}. The results of the analyses 
are given in \S~\ref{sec:result}. 
A final discussion of the results is given  in \S~\ref{sec:disc}.

\begin{table}
  \begin{center}
    \begin{tabular}{ccccc}
      \hline
      Model & $\Omega_m$ & $\Omega_\Lambda$ & h & $\sigma_8$ \\ 
      \hline
      \LCDM & 0.3 & 0.7 & 0.7 & 0.9 \\
      OCDM  & 0.3 & 0 & 0.7 & 0.9 \\
      SCDM  & 1.0 & 0 & 0.5 & 0.7 \\
      \hline
    \end{tabular}
    \caption{Cosmological parameters used for the differents CDM
      models}
    \label{tabmodels}
  \end{center}
\end{table}

\section{Simulations}
\label{sec:sim}

W05 has analyzed one and only one particular patch of the universe,
namely the LV that extends around the LG. The LG is not a unique or an
unusual object in the universe, yet it has its own characteristics that
affect the outcome of any dynamical test that would 
be applied to it. Our
main goal  is to check the validity of the W05 approach, in
the context of the environment of the LG.  This is to be achieved by
applying the W05 analysis to LG-like objects found in appropriate
N-body simulations. The key to a successful study is to properly select
LG-like objects from the simulation and analyze the LV around these
objects (see \S~\ref{sec:obj}). Yet the selection of the LG-like objects
is based on the properties of the LG itself and not on its environment
within the LV. This has encouraged us to use CSs as a 'laboratory' for
testing the W05 procedure in conditions very similar to the ones
prevail in the LV and its immediate surrounding. In addition we have
used unconstrained simulations for reference.  The comparison of the
results obtained in the constrained and unconstrained simulations will
shed light on the question of whether the findings of W05 are naturally
expected in CDM dominatted cosmogonies.

The data used to constrain the initial conditions of the simulations is
made of two kinds. The first data set is made of radial velocities of
galaxies drawn from the MARK III \citep{mark}, SBF \citep{sbf} and
the \cite{kar05} catalogs. Peculiar velocities are less affected by
non-linear effects and are used as constraints as if they were linear
quantities \citep{zar99}. This follows the CSs
performed by \cite{kra02} and \cite{kly03}. The other constraints are
obtained from the catalog of nearby X-ray selected clusters of galaxies
\citep{rei02}. Given the virial parameters of a cluster and assuming
the spherical top-hat model one can derive the linear overdensity of
the cluster. The estimated linear overdensity is imposed on the mass
scale of the cluster as a constraint. Different CSs with different
random realizations have been calculated and they all exhibit a clear
and unambiguous LSC-like structure that dominates the entire
simulation, much in the same way as in the actual universe in which the
LSC dominates the nearby LSS. The simulations do vary with respect to
the particular details of the LG-like object that is formed roughly in
its actual position. All simulations used here are based on the same
random number realization  of the initial conditions.

Five simulations have been performed so far. Three constrained ones have
been performed within the framework of the \LCDM, OCDM and SCDM models.
In addition,  unconstrained simulations of the \LCDM and OCDM cosmologies
have been performed for the sake of comparison and benchmarking.  
Table \ref{tabmodels}
presents the cosmological parameters 
of the different simulations.
All simulations  correspond to a periodic  cubic box of  64 h$^{-1}$ Mpc  on a
side.  We made a random realization of the corresponding  power
spectrum for each cosmological model with a large number of particles 
($2048^3$). 
At the same time, a constrained realization of the density
field for the same power spectra was done in  an eulerian mesh  of
$256^3$ grid points. After fourier transforming  both constrained and
unconstrained  density fields, we substitute  the  fourier modes  of
the constrained  field  into  the unconstrained one. Finally, we used
Zeldovich approximation  to compute the 3D displacement field for an
initial redshift of z=60.  Once we got the displacements for a refined
mesh  of $2048^3$ grid points, we  used it to estimate the initial
conditions for a resampling  of  $256^3$ dark matter particles in
total. In this way, we are able to zoom into a particular area of the
simulations and resimulate them with much higher resolution, 
up to the maximum resolution possible $(2048^3)$, 
 having the same structures as in the low resolution simulations 
(see  \citet{kly01} for a more detailed information about zoomed
simulation techniques). Thus,  
all the numerical experiments that are reported here have the
same number of particles ($256^3$). This translates into a mass per
particle of $1.3 \times  10^7 h^{-1}  M_\odot$ for \LCDM and OCDM  and  $4.3
\times   10^7 h^{-1} M_\odot $ for SCDM simulations.

We have used the parallel TREEPM N-body code  GADGET2 (\cite{gadget2}) 
to run these simulations. An  uniform mesh of $512^3$  grid points was
used to compute the long-range gravitational force by means of the
Particle-Mesh algorithm. A constant comoving Plummer equivalent 
 gravitational smoothing scale of 20 $h^{-1}$ kpc was set  at high
 redshift and we changed it to  5 $h^{-1}$  kpc physical scale 
since z=3 till z=0.  The number of timesteps to complete the evolution
from z=60 till z=0 ranges from 5000 to 7000 depending on the
simulation.   We employed a variety of  parallel computer architectures
(SGI-ALTIX, IBM-SP4, Opteron-clusters) during
the course of this work. Using  16 processors simultaneously, we
completed  one run in  about 2 cpu days. 
   
In what follows, we will  use the name    \LCDM, OCDM and SCDM 
 for the   simulations  with Constrained Initial conditions in
the different cosmological models. The names {\LCDM}u and OCDMu 
will refer  to the  two different unconstrained realizations 
 in the \LCDM\ and OCDM models respectively.
As an example  of how the simulations look like, we show  
in Figure \ref{boxL} a projection of the dark matter distribution in
 the  \LCDM\ simulation box at z=0.

\section{Selection of LG  Candidates}
\label{sec:obj}

\begin{table}
\begin{tabular}{|ll|c|}
\hline \hline
\multirow{4}{*}{Components} & Kind & MW + M31  \\ 
 & Mass & $125 \leq V_c \leq 270$ km/s \\ 
 & Separation & $s \leq 1$ Mpc/h \\ 
 & Relative velocity & $V_r < 0$ \\ 
\hline
\multirow{2}{*}{$\nexists$ neighbours} & Distance to LG & $d_{neigh} < 3$ Mpc/h \\ 
 & Mass & $V_c \geq V_{c, comp}$ \\ 
\hline
\multirow{2}{*}{Virgo halos} & Distance to LG & $5 \leq d_{Virgo} \leq 12$ Mpc/h \\ 
 &  Mass & $500 \leq V_c \leq 1500$ km/s\\ 
\hline \hline
\end{tabular}
\caption{Constrains used to find LG candidates
 following the Macci\`{o} \etal (2005) criterion. 
The circular velocity has been used for the mass contrains.}
\label{LGc}
\end{table}

Dark matter  halos were found in simulations using two object finding
methods: The Bound Density Maxima (BDM) algorithm (\cite{bdm}) is  
based on finding local center of mass  in spheres of variable radius
starting from randomly selected particles in the simulation. 
 The Amiga Halo Finder (\cite{amiga}), on the contrary, finds local
density  maxima  from an adaptive mesh hierachy. 
In both cases, an iterative procedure to find local centre of mass from
density maxima is used. Particles  that are not gravitatonally bound to
the halo potential are also removed.  We took halos composed by  more than
100 dark matter particles, which translates into a minimum mass per halo of
 $M_{min}= 1.3 \times  10^9 h^{-1}  M_\odot $ for the \LCDM\  and OCDM simulations 
 and $M_{min}= 4.3 \times  10^9 h^{-1} M_\odot $ for the SCDM simulation. 
We indentified the same objects with both methods. For the work
reported here we have used the halo catalogues obtained  by the public
available AMIGA halo finder code 
(http://www.aip.de/People/AKnebe/AMIGA/).

\begin{figure}
\resizebox{\linewidth}{!}{\includegraphics{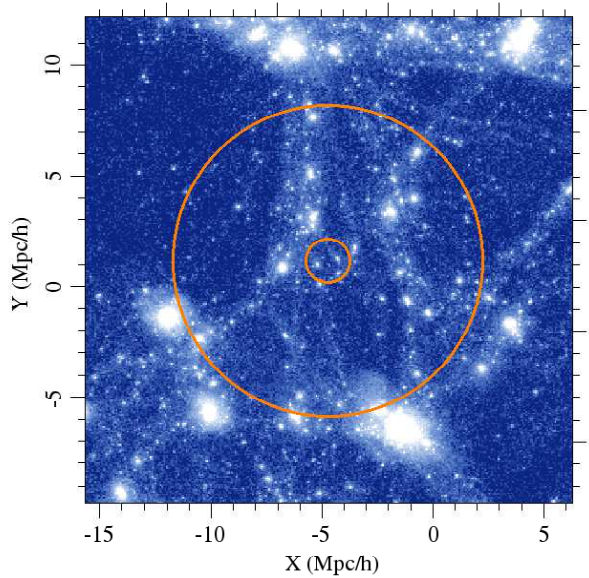}}
\caption{Projected dark matter distribution around  the  best LG candidate
  in the \LCDM\ simulation in supergalactic coordinates.  The outer
circle delimits  the Local Volume and the inner circle represents
the LG position  }
\label{spec_c}
\end{figure}

To  identify LG candidates from the halo distribution,
 we selected those objects that  fulfill the strict requirements 
as given in \cite{gov97} and
 \cite{mac05}. They are summarized in  Table \ref{LGc}. In brief, 
we searched for two  halos similar to Milky Way and M31 galaxies, 
without neighbours  with  masses as high as any  of the LG members 
and with a Virgo-like halo at  an appropiate distance.  
A few tens of LG-like objects have been
 identified in each simulation ($23$ objects in \LCDM, 34 in $\Lambda$CDMu, $41$
 in OCDM, $58$ in OCDMu and 37 in SCDM).   One of the LG-like object
 for the \LCDM simulation  which closely resembles the actual LG in
 terms of  its mass and position, is presented in Figure \ref{spec_c}.

\begin{figure}
\resizebox{\linewidth}{!}{\includegraphics{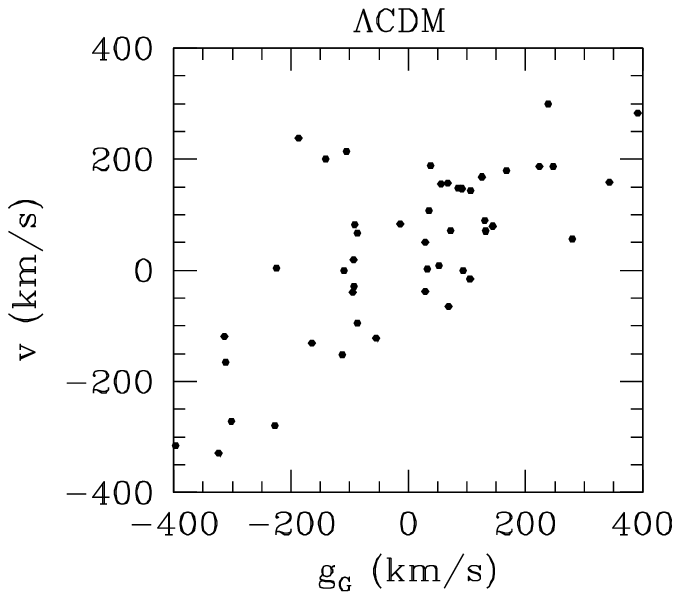}\includegraphics{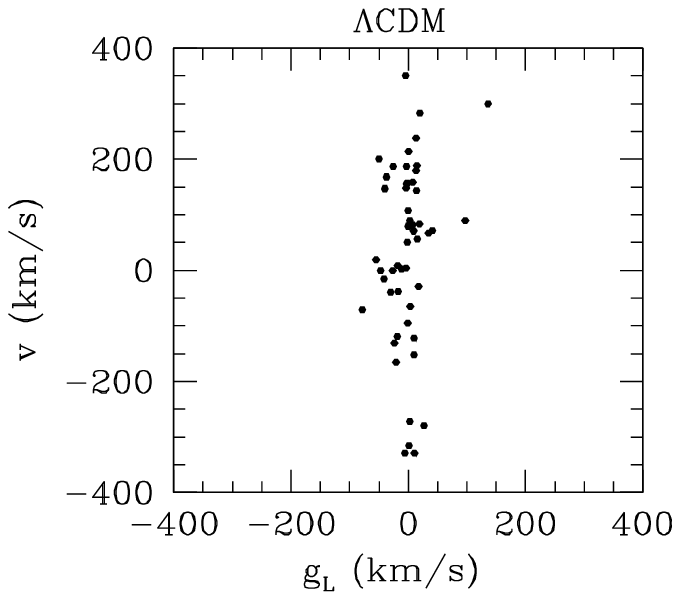}}
\caption{Plots of $v$ \vs $g_G$ and $v$ \vs $g_L$ for the candidate of Figure \ref{spec_c}.
}
\label{spec}
\end{figure}

\section{Analysis}
\label{sec:analysis}

In order to study the dynamics of the LG-like objects found in the
simulations, we have first computed the local Hubble flow 
in spheres of $7\hmpc$ around each candidate, 
and have also estimated the local overdensity within these spheres 
from the total mass inside them.  All  DM halos
within the LV around each LG-like objects have been identified and the
gravitational field acting on each halo has been calculated in two
different ways.  First, the local gravitational field field is
calculated like in W05 by the summation over the pairwise newtonian 
interaction. Namely, the field acting on the i-th halo is given by \be
\tilde{\bg}{^{tot}_{l,i}} = -G \sum_{j\neq i} M_j \frac{ {\bf r}_j- {\bf r}_i
}{[( {\bf r}_j-{\bf r}_i )^2+A^2]^{3/2}} {\bf \cdot \hat{r}}_i,
\label{eqloc}
\ee where,  following W05,  a softening parameter of $A=1.2\hmpc$ is
introduced.  Then we calculate also the 'true' gravitational field,
namely the field calculated by the N-body code of the full mass
distribution in the computational box. This is defined as the global
gravitational field, (see the appendix for further information).

As in  W05,  the local gravitational field is decomposed into two
terms, the contribution of a smooth background of matter and a
fluctuating part given by the point mass distributions.  Since we are
interested in deviations from the average, we have to substract the
linear term contributed by the background, $\bg_{bg}$, from the local
gravitation field 
\be 
\tilde{\bg}_l = \tilde{\bg}{^{tot}_l} - \tilde{\bg}_{bg}. 
 \ee The
calculation of  the background term is done in two ways. First,
following W05, the background solution is fitted by a linear  term in
${\bf r}$.  Alternatively, the background solution is calculated by the
exact solution for the unperturbed universe: 
\be 
\tilde{\bg}_{bg} = -G \frac{4  \pi}{3} \Omega _m \rho_c {\bf r} 
\ee 
The two methods give virtually identical
results and the fitting method has been used here, so as to be
consistent with analysis of W05. 
We have also substracted the anisotropic background estimated
  from the tidal field (see \S{} \ref{sec:a-v}).

As we have mentioned earlier, the purpose of the present study 
is to compare gravitational accelerations and velocities. 
To facilitate such a comparison the gravitational field is 
scaled by the linear theory prediction, 
\be
\bg_x=\frac{2f(\Omega_m,\Omega_\Lambda)}{3H_0\Omega_m}\ \tilde{\bg}_x 
\ee 
where $x$ stands
here for the local or global field.  The velocity-gravity scaling
factor  is given by 
\be f(\Omega_m,\Omega_\Lambda) \approx
\Omega_m^{0.6}+\frac{1}{70}\Omega_\Lambda\left(1+\frac{1}{2}\Omega_m \right)
 \ee
(e.g. \cite{pee80}, \cite{lah91}).  Throughout the paper the gravitational
fields will be  represented by their  scaled versions.

Our analysis is exemplified by Figure \ref{spec_c} which shows
one of the LG-like objects in the \LCDM\ constrained simulation.
 The figure shows the  matter distribution in a box of 
$20 \hmpc$ centered on the simulated LG
and projected on the Supergalactic plane. In Figure \ref{spec}, two
scatter plots show the scatter of the local and global gravitational
accelerations, 
 \vs the peculiar velocity of all DM halos in this
particular LV. In what follows, all quantities shown correspond to
 projections along the  line of sight with the observer located  in the
 center of mass of the LG candidates.

\begin{figure*}
\resizebox{\textwidth}{!}{ \includegraphics{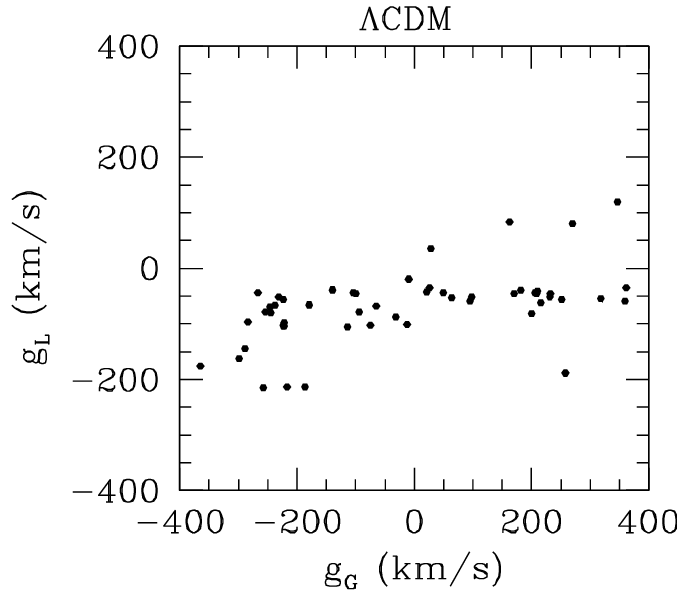}\includegraphics{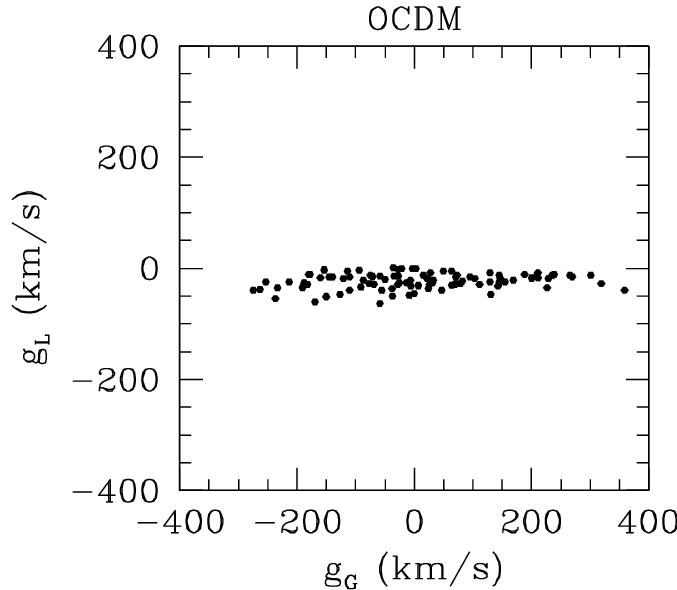}\includegraphics{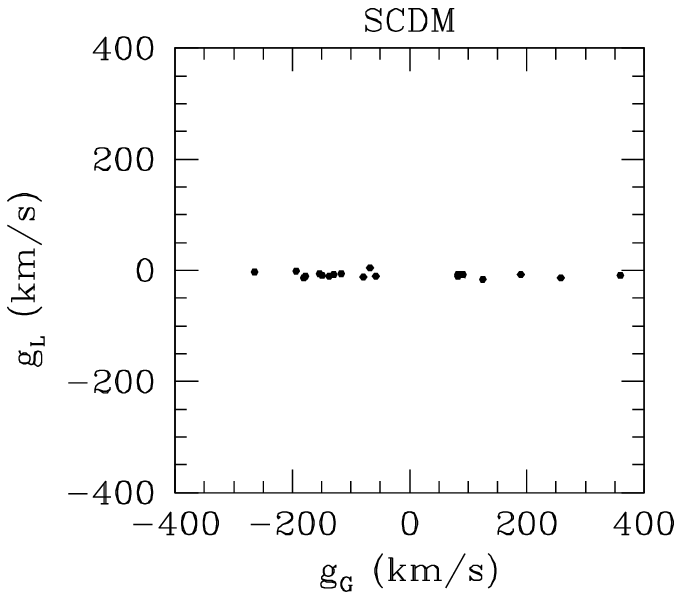}}
\caption {
Local \vs global scaled acceleration for one of the LG
 candidates in  the  constrained \LCDM, OCDM and SCDM simulations respectively.
}
\label{loc_g}
\end{figure*}

\begin{figure*}
\resizebox{\textwidth}{!}{\includegraphics{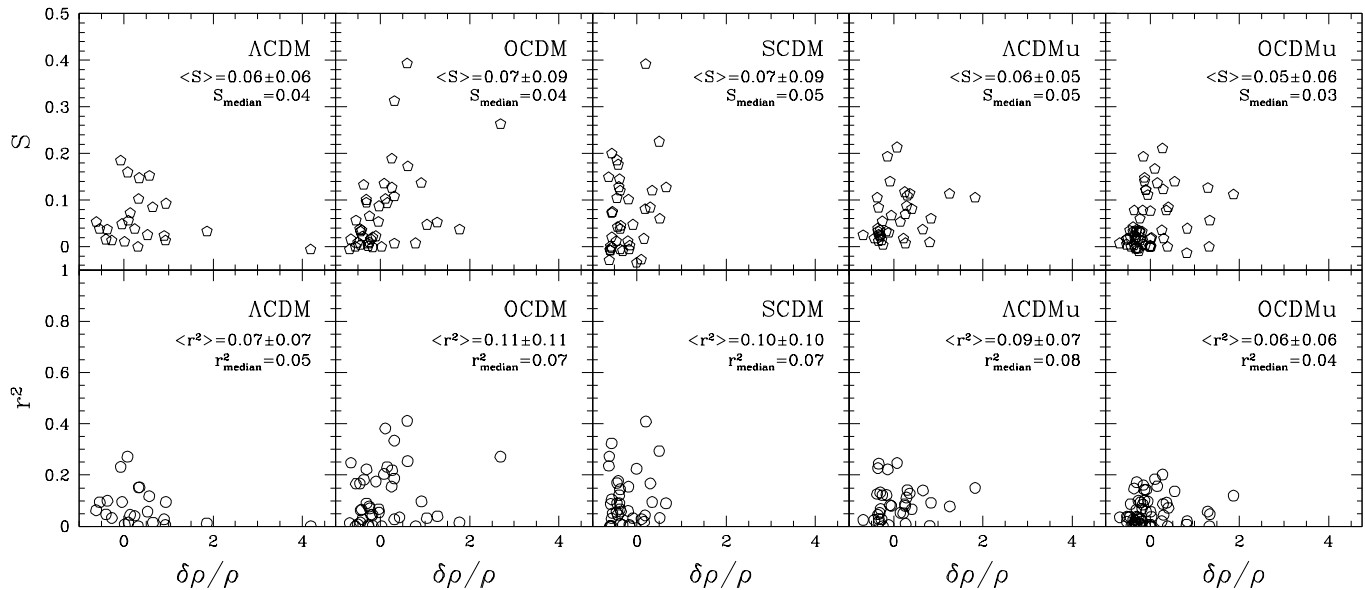}}
\caption {Slopes (first row) and Pearson's  correlation coefficient 
(second row) of the linear fit  of $g_L$ \vs $g_G$  as a function of
overdensity  calculated within the simulated Local Volumes for all
candidates found in  each  simulation. The median  and  the average 
slope  and   correlation coefficient with their one sigma errors 
 for all LGs in each simulation are also shown. 
}
\label{loc_g_cor}
\end{figure*}

\section{Results}
\label{sec:result}

\subsection{Relation between local and global accelerations}
\label{sec:loc-glob}

Figure \ref{loc_g} shows  scatter plots
of  local \vs global gravitational fields  for some of the  Local
Volume candidates extracted from simulations.  In order to study the
relation between both fields,   a linear fit to the $g_L$ \vs $g_G$ distribution has been made
for  all the LG candidates  found in the different  simulations.

Figure \ref{loc_g_cor} shows the  least square fit  slopes and
correlation coefficients  of the local to the global gravitational fields
as a function of  the    overdensidity $\delta \rho / \rho$ measured within each
of the simulated LV's.  We find that the local and global field are 
uncorrelated, showing  an extremely small correlation coefficient and a
 mean and median slope of roughly $0.1$ rather than unity.

 We have tested  the effect of taking only  haloes with high mass
 (more than $10^{11}\hmsun$).
 The results are  very  similar to the previous ones,  because the
 lightest haloes  do not contribute much to the gravitational field.
The overdensity was calculated by computing all matter inside the local
volumes. When using only mass in halos to estimate the overdensity, 
results do not change at all.

\begin{figure}

\resizebox{\linewidth}{!}{\includegraphics{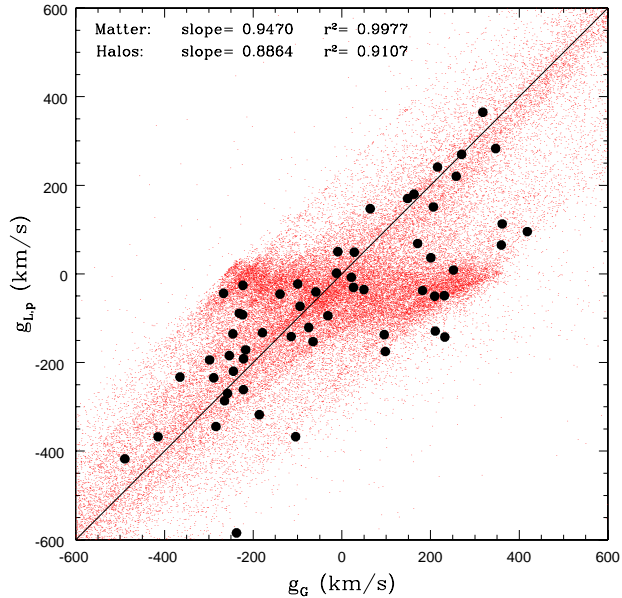}}
\caption { Local \vs\ global gravitational accelerations computed 
 taking into account all the   particles within the LV
 for the same  candidate in the \LCDM\ simulation as in  Figure
  \ref{loc_g}. The small points represent the individual dark matter
  particles. Thick solid points correspond to halos in which
  accelerations have been computed by the average of all particles
  inside them. The straight line shows the equivalence between both accelerations.}
\label{all_partic}
\end{figure}

\begin{figure}
\resizebox{\linewidth}{!}{\includegraphics{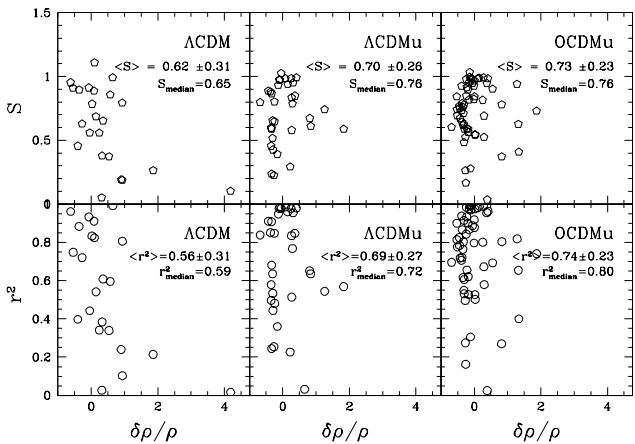}}
\caption{Slopes (first row) and Pearson's  correlation
 coefficients (second row) for the  $g_L$ \vs $g_G$ least square
 fits  for  all the LG
 candidates found in the \LCDM\ and OCDM   simulations as a function of
 overdensity.   All  particles within  simulated  LVs have been taken
 in   the computation of the local accelerations. 
A gravitational smoothing parameter of 5 $h^{-1}$ kpc is
 assumed to compute the newtonian pairwise forces between particles.  
}
\label{fig7}
\end{figure}

In the preceding analysis the local gravitational field is calculated
by summing over the pairs of DM halos, assuming they carry all the mass
in the LV
and that their mass is spread out over a relatively large area
described by the $A$ smoothing parameter.
The poor correlation between
the local and global gravitational fields has led us to relax 
the assumption that the mass is traced by the DM halos and calculate 
the local field contributed by all dark matter particles in the LV. 
This is presented in Figure \ref{all_partic},
which shows a very tight correlation between $\bg_L$ and $\bg_G$ for
one particular LG-like object in the \LCDM\ CS.
A better correlation  with  a slope very close to unity  are 
 obtained when we  include  all the  particles that
 constitute the inter-halo medium, rather than by the intra-halo
 particles only, as can be seen in Figure \ref{fig7}.
 Moreover, the only gravitational smoothing  done in this case
correspond to that included in the simulation, which is of the order of
kiloparsec scale, contrary to the strong smoothing of 1.2 Mpc used in
W05 analysis and in our previous estimates.
The horizontal branch around the origin  observed in Figure \ref{all_partic} is
  mainly due to particles not bound to halos. 
These particles are more affected by the external field than particles bounds to halos, and therefore for these the local acceleration is much smaller that the global one.

Obviously, the correlation between the local and global acceleration improves as the LV increases. 
This has been checked  for the LG-like object of  Figure \ref{all_partic}. Assuming a LV of 
a 30 $\hmpc$ radius  the correlation improves to $S=0.47$ and $r^2=0.58$. However this is still
considerably worse than the  $S=0.89$ and $r^2=0.91$ obtained by considering all the particles
within the 7 $\hmpc$ LV (Figure \ref{all_partic}).

 The lack of correlation between  the local and global gravitational 
accelerations, strongly leads to expect the absence of correlation
 between the local gravitational field and the peculiar
 velocities. This is clearly confirmed by the analysis presented in 
 \S{} \ref{sec:a-v}.

\subsection{Accelerations and velocities}
\label{sec:a-v}

\begin{figure*}
\resizebox{\textwidth}{!}{ \includegraphics{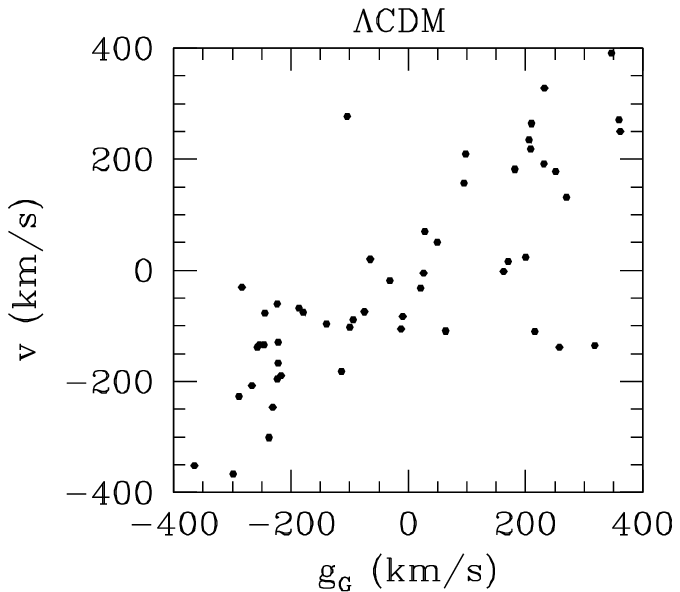}\includegraphics{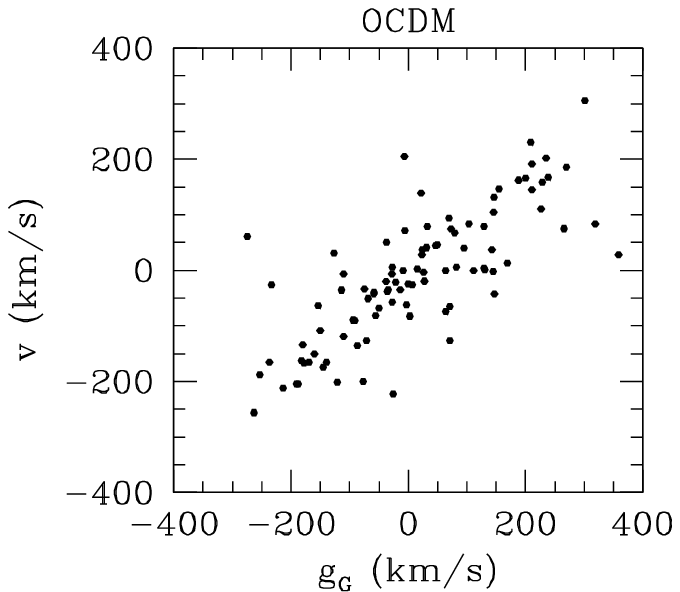}\includegraphics{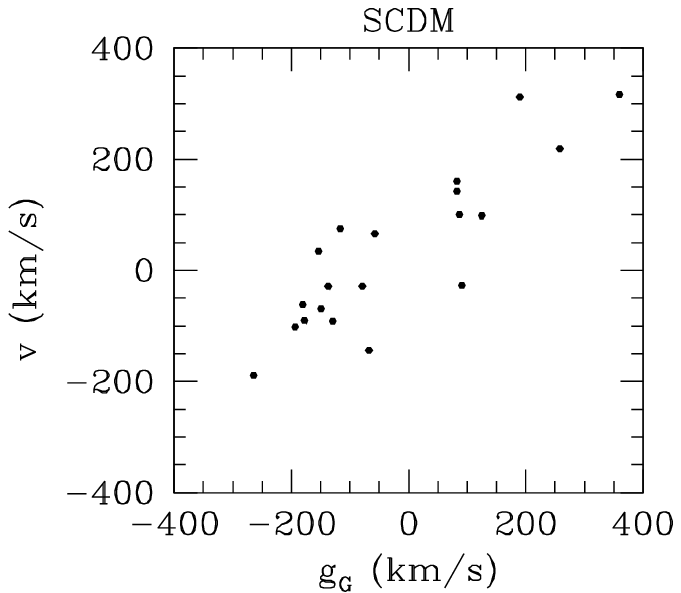}}
\caption {
Peculiar velocity \vs global scaled acceleration of halos inside  the
LV for the same candidates as in  Figure \ref{loc_g}.
}
\label{v_g}
\end{figure*}

\begin{figure*}
\resizebox{\textwidth}{!}{\includegraphics{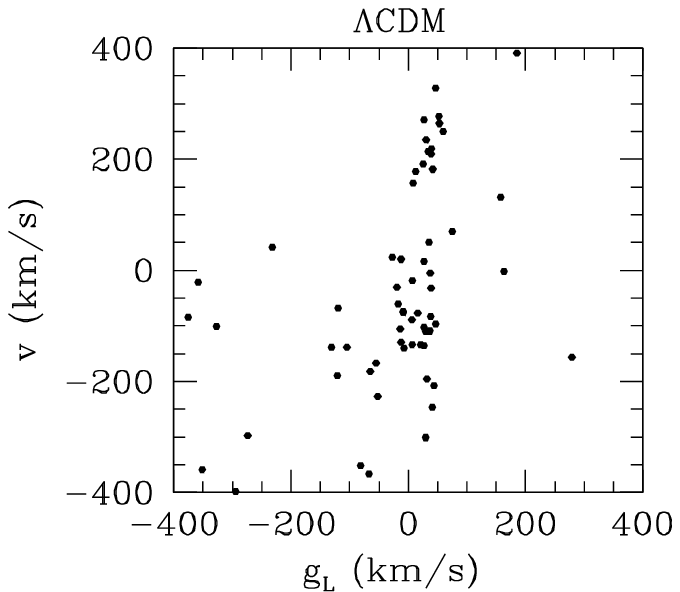}\includegraphics{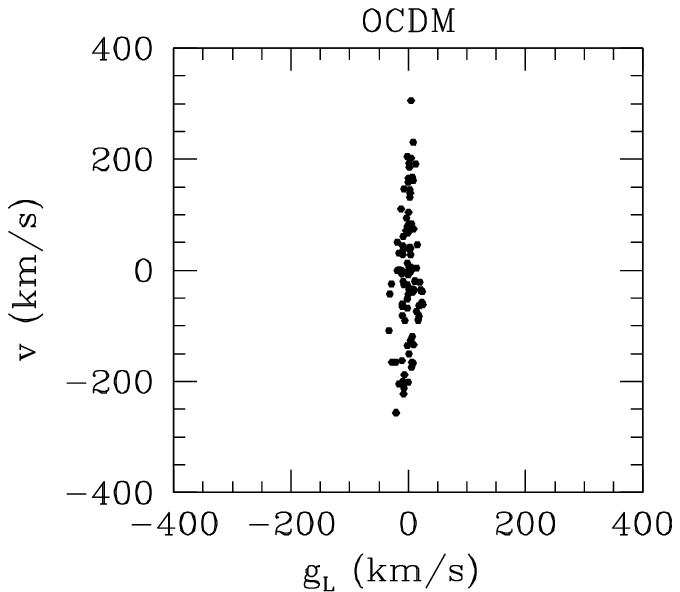}\includegraphics{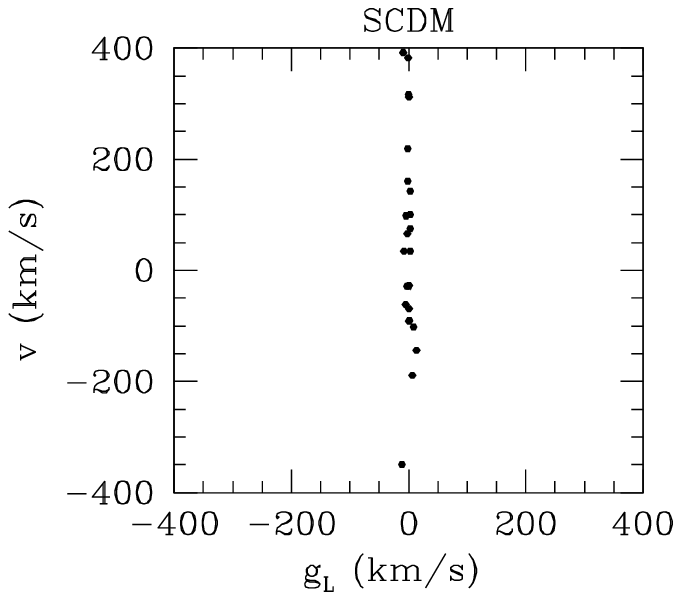}}
\caption {
Peculiar velocity \vs local scaled acceleration for halos within the LV
for the same candidates as in  Figure \ref{loc_g}.
}
\label{v_loc}
\end{figure*}

\begin{figure*}
\resizebox{\textwidth}{!}{\includegraphics{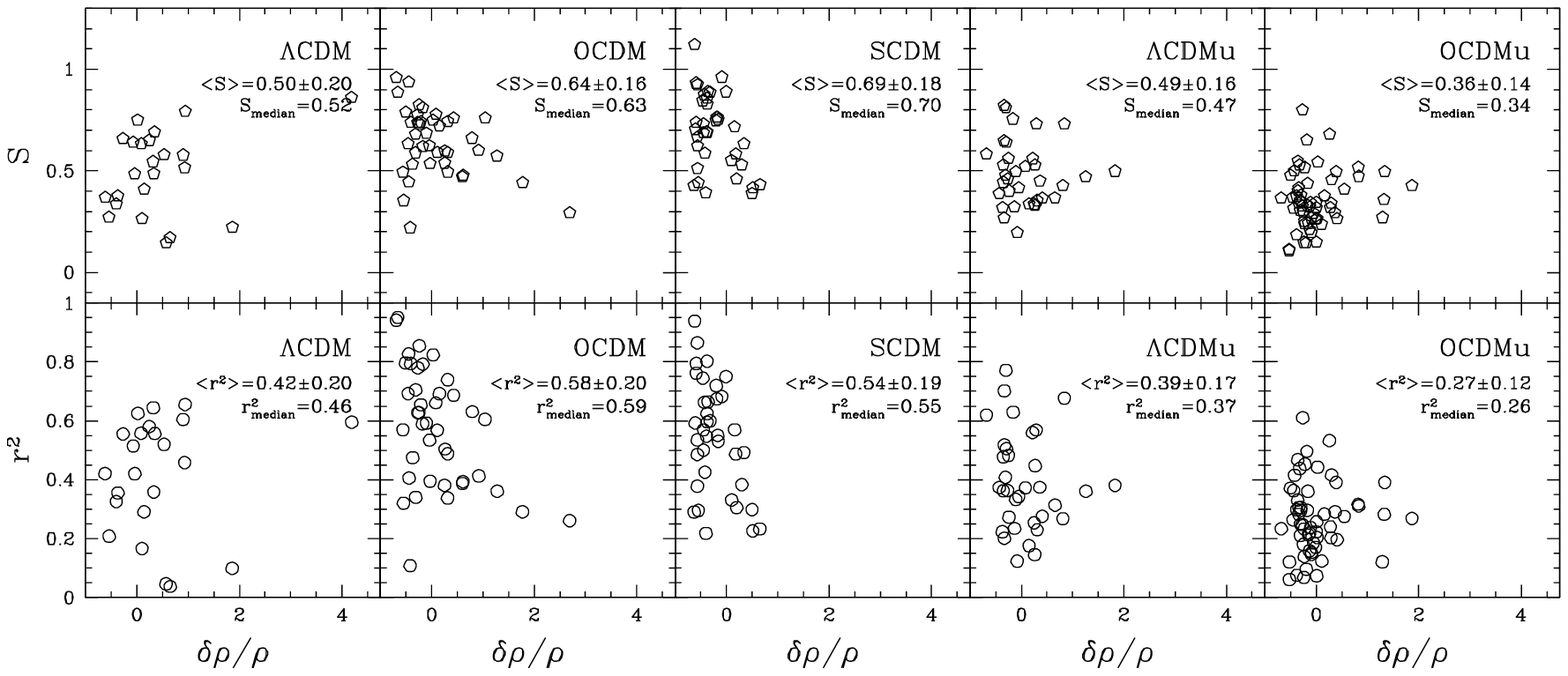}}
\caption {Same as Figure \ref{loc_g_cor} but for the  peculiar velocity
  \vs global scaled aceleration  fits.
}
\label{v_g_cor}
\end{figure*}

\begin{figure*}
\resizebox{\textwidth}{!}{\includegraphics{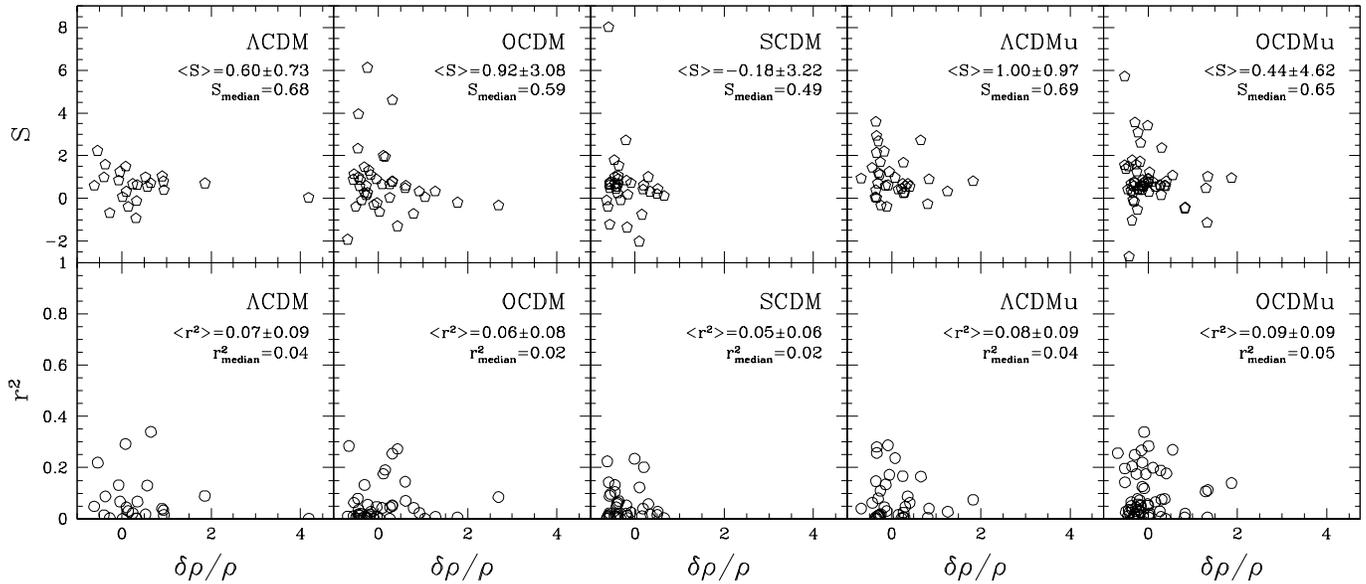}}
\caption {
The same as Figure \ref{loc_g_cor} but for the  peculiar velocity \vs local scaled acceleration fit.
}
\label{v_loc_cor}
\end{figure*}

Figures \ref{v_g} and  \ref{v_loc}  present the peculiar velocities as a
function of the scaled global and local   gravitational accelerations
around some selected LG-like objects in the different simulations.  The
slopes and the correlation coefficients of the linear fits for the
relation between the peculiar velocities and the gravitational
accelerations acting on halos in the LV around all LG-like objects are
given in Figures \ref{v_g_cor} and \ref{v_loc_cor}.  The least square fit
analysis confirms the visual impression of Figures \ref{v_g} and
\ref{v_loc}. The peculiar velocity of halos in the LV of LG-like
objects is clearly correlated with the global acceleration but with a
non-negligible scatter, with a mean and median correlation
coefficient  around $0.5$ for the constrained simulations 
and somewhat lower for  the
unconstrained ones (Figure \ref{v_g_cor}). 
The mean and the median slope of the
linear relation between the velocity and scaled global acceleration is
in the range of $0.4$ to $0.6$.  The distribution of the fitted slopes
shows some dependence on the mean overdensity ($\delta \rho /\rho$) in the LV. The
width of the distribution decreases with the overdensity. At low $\delta \rho
/\rho$'s the slope ranges from roughly $0.2$ to almost unity but at $\delta \rho
/\rho \gtrsim 1$ the slope shows a narrow scatter around its mean value of $\approx
0.5$.  Upon scaling of the gravitational field the linear theory
predicts the slop to be unity. It follows that the amplitude of the
peculiar velocities is smaller than what is expected by the linear
theory.

The peculiar velocities do not show any correlation with the local
accelerations (Figures \ref{v_loc} and \ref{v_loc_cor}). Their linear
fit yields extremely low correlation coefficients and the fitted slope
has no meaning. One should recall here that this lack of correlation
has already been anticipated from the lack of correlation we found 
between the local and global accelerations (\S\ \ref{sec:loc-glob}).

The calculation of the local gravitational acceleration neglects the
contribution of the tidal field, which is induced by the inhomogenous
matter distribution outside of the LV.  Obviously, for objects like the
LG, with the Virgo cluster located just outside of the LV, the tidal
field cannot be neglected.  The local gravity-velocity
correlation should be improved by adding the tidal field into the
fitting procedure. Indeed, we follow W05 and extend the fitting
procedure to
\begin{equation}
  \sigma^2=\frac{1}{N}\sum (g_{L}-v+{\bf v}_0 \cdot {\bf \hat{r}}+{\bf \hat{r}\cdot H \cdot r})^2,
  \label{sig2}
\end{equation}
where ${\bf v}_0$ is an unknown vector and ${\bf H}$ is a symmetrical
tensor with six unknown quatities. The nine free parameters are found
by minimizing the scatter. Note that in the limit of the linear theory
and for a small LV ${\bf v}_0$ should be set to zero and the symmetric
tensor $H$ should be traceless. As the above assumptions do not hold
for the LG we allow for a finite ${\bf v}_0$ and for $H$ to have a
trace.

Figure  \ref{v_loc_cor_m} shows that indeed adding the
${\bf v}_0$ and ${\bf H}$ terms improves somewhat the correlation
between the local gravity and peculiar velocities. Yet, they are weakly
correlated with $r^2 \approx (0.1 \ - \ 0.2)$ for most LG-like objects. The
minor improvement is not surprising. The Virgo cluster  is located at a
distance of about $10 \hmpc$ and the LV is defined by a sphere of
radius of $7\hmpc$. Hence, modeling the tidal field by a spatial linear
expansion, as is implicitly assumed in Eq. \ref{sig2} constitutes a
poor fit of the tidal field.

\section{Discussion}
\label{sec:disc}

We have clearly demonstrated in this paper 
 that the W05's analysis is not 
 expected
  to yield a simple 
and clear correlation between the gravitational field
calculated from the mass distribution within the local volume and the
peculiar velocities of halos within that volume.  This result
invalidates W05's basic assumption that the peculiar velocity field
traces the gravitational field in a simple fashion, and therefore the
statements concerning the role of the gravitational instability in the
LV are 
not valid.

Setting aside issues concerning the practical limitations posed by
observations of the LV and the uncertainty in estimating the dynamical
mass of luminous galaxies there are three main theoretical reasons why
W05's analysis fails. The most obvious one is that the gravitational
field is assumed to be traced by the galaxies, treated as point-like
particles. Figs. \ref{loc_g} and \ref{loc_g_cor} show the poor
correlation between the actual gravitational field and the local field
induced by the galaxies. Now, this poor correlation is mostly due to
the sampling of the field by the galaxies, as manifested by Fig.
\ref{all_partic} which shows a clear correlation between the global
field and the local field that is induced by all dark matter particles 
in the LV.

\begin{figure*}
\resizebox{\textwidth}{!}{\includegraphics{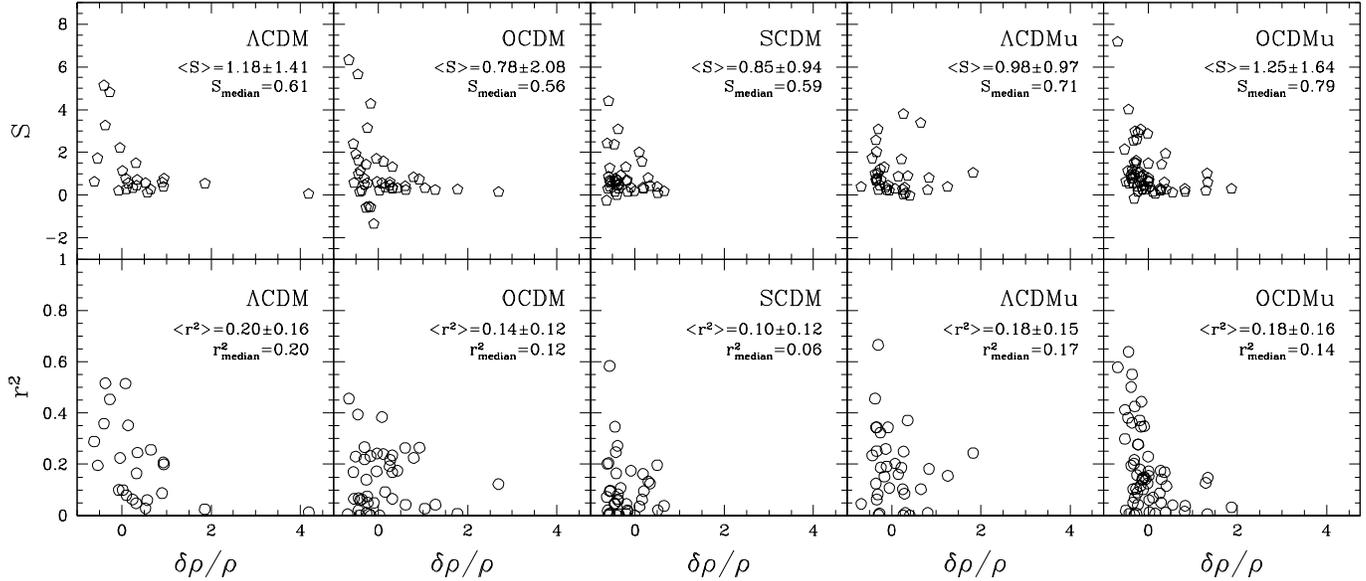}}
\caption {
Same as  Figure \ref{v_loc_cor},  but including the linear tidal field
  into the fitting procedure (see Eq. \ref {sig2})}
\label{v_loc_cor_m}
\end{figure*}

The other reasons for the breakdown of the simple gravity - velocity
relation of the linear theory are both related to the tidal field,
hence the sheer of the velocity field. The simpler reason is that in
solving the Poisson equation one should not neglect the homogenous
solution, namely the tidal field. Now, in principle this can be easily
corrected by adding a (spatial) linear term to the gravitational field
that scales with the traceless shear tensor (Eq. \ref{sig2} and Figure
 \ref{v_loc_cor_m}).  Thus by adding six free
parameters to the fitting procedure one might be able to account for
the tidal field. However, the size of the LV is such that the spatial
linear expansion of the tidal field would fail and lead to an incorrect
estimation of the gravitational field. For the LV centered on the LG,
the tidal effect of the LSC cannot be represented by a linear term.

The other reason for the inadequacy of the linear theory is more
subtle. It has been shown that in the quasi-linear regime the growth of
the density contrast depends on the magnitude of the shear tensor
\citep{hof86,hof89,zar93,van94,ber94}.
The shear dependence introduces a non-local term in the equations that
govern the growth of structure in the quasi-linear regime.  Indeed,
Fig.  \ref{v_g} shows a tight linear relation between the peculiar
velocities and the scaled global gravitation field. Yet, the constant
of proportionality is less than unity for all the LG-like objects in
all the simulation (except of one single object) as is predicted by the
linear theory prediction (in agreement with \citet{hof89}). This
behavior implies that under the optimal conditions of a full knowledge
of the gravitational field a linear relation between the gravity and
velocity field is expected within the LV around LG-like objects.
However, using the slope of the relation as a way of measuring $\Omega_m$
would underestimate its true value.

W05 attempted to find a simple linear relation between the peculiar
velocities and local gravitational field of galaxies in the LV. A
careful analysis of the dynamics within the LV around LG-like objects
identified in constrained simulations of the local universe and in
unconstrained simulations in flat-\LCDM, OCDM and flat-matter only CDM
cosmologies shows that a lack of correlation is to be expected.  Hence
we cannot support the claim that {\em 'either dark matter is not distributed
in the same way as luminous matter in this region, or peculiar
velocities are not due to fluctuations in mass.'}

\section{Acknowledgements}
We appreciate very much  the comments and discussions with Yago
Ascasibar, Anatoly Klypin, Andrey Kravtsov, Stefan Gottl\"ober and  
Yaniv Dover. We would like to thank Astrophysikalisches Institut Potsdam for
allowing us to use the Sansoucci supercomputer opteron clusters and to
host us several times during the course of this work.  
We also thank CIEMAT (Spain)
to allow us to use their SGI-ALTIX supercomputer  and to NIC J\"ulich
(Germany) for the  access to the IBM-Regatta p690+ JUMP supercomputer.
GY would like to thank also MCyT for financial support under 
 project numbers  AYA2003-07468  and BFM2003-01266. 
LAMV acknowledges financial support from  MCyT (Spain) under project
BFM2003-01266 and from Comunidad de Madrid through a  PhD fellowship.
YH acknowledges the support of  ISF-143/02, the Sheinborn
Foundation and the DFG for a Mercator Gastprofessur at
Potsdam University.

\section{Appendix: The GADGET estimation  of gravitational   accelerations}

Throughout the paper the global and local gravitational accelerations
have been   compared.
The local  accelerations within the LV  were 
 calculated by summing the Newtonian pairwise field.
This can be repeated for the global field by summing over all the
particles of the simulation. This direct sum is computationally very
costly and does not take into account the contribution of the infinite
periodic boxes.  We used instead the particle accelerations calculated
by the TREE-PM algorithm by the GADGET code.  The relation between 
the GADGET acceleration with $\bg_G$ is as follows:

The physical  $\textbf{r}$ and comoving $\textbf{x}$ coordinates are related by:
$\textbf{r} = a\textbf{x}$. The global gravitational field equals
 the physical acceleration of an object 
$\ddot{\textbf{r} }= \bg$.

Now,  the GADGET code provides an acceleration-like term defined as:
\begin{equation}
\bff_p=\frac{1}{a} \frac{d}{dt}\left( a\cdot \bv_p\right)=\dot{\bv}_p + H\cdot \bv_p,
\label{fpdef}
\end{equation}
where $\bv_p$ is the peculiar velocity and $a$ is the expansion scale factor. 

It follows that 
\begin{equation}
\ddot \br = \bff_p + \br \frac{\ddot a}{a}.
\label{r_f_a}
\end{equation}
Recalling that $\frac{\ddot a}{a}=-\frac{4\pi G\rho_c\Omega_0}{3}$ one gets at the end
\begin{equation}
\bg_G = \bff_p - \frac{1}{2}H^2\Omega_0 \cdot \br.
\end{equation}
This is the value we used for the  global acceleration of each dark
matter particle. The total acceleration of halos was computed by
averaging this quantity  over the entire number of particles belonging
to each halo.

\end{document}